# Pauli Spin Blockade of Heavy Holes in a Silicon Double Quantum Dot

Ruoyu Li,[†] Fay E. Hudson,[‡,¶] Andrew S. Dzurak,[‡,¶] and Alexander R. Hamilton[*,†]

[†]*School of Physics, University of New South Wales, Sydney NSW 2052, Australia*

[‡]*Australian National Fabrication Facility, University of New South Wales, Sydney NSW 2052, Australia*

[¶]*Centre of Excellence for Quantum Computation and Communication Technology, School of Electrical Engineering and Telecommunications, University of New South Wales, Sydney NSW 2052, Australia*

E-mail: Alex.Hamilton@unsw.edu.au

## Abstract

In this work, we study hole transport in a planar silicon metal-oxide-semiconductor based double quantum dot. We demonstrate Pauli spin blockade in the few hole regime and map the spin relaxation induced leakage current as a function of inter-dot level spacing and magnetic field. With varied inter-dot tunnel coupling we can identify different dominant spin relaxation mechanisms. Applying a strong out-of-plane magnetic field causes an avoided singlet-triplet level crossing, from which the heavy hole $g$-factor $\sim 0.93$, and the strength of spin-orbit interaction $\sim 110$ $\mu$eV, can be obtained. The demonstrated strong spin-orbit interaction of heavy hole promises fast local spin manipulation using only electrical fields, which is of great interest for quantum information processing.

# Keywords

There is great interest in developing solid-state quantum dot based spin-qubits for quantum information applications.[1] Early work on electron spins in III-V semiconductors has resulted in the demonstration of single spin isolation,[2] manipulation,[3] and readout.[4] The use of semiconductors with strong spin-orbit interactions (SOI) has allowed rapid local control of individual spins using only electric fields.[5] However, in III-V materials, the ever-present hyperfine interaction between electrons and the nuclear spins in the host crystal limits the spin dephasing time $T_2^*$.[3] In contrast, electron spin qubits in silicon quantum dots, especially in isotopically enriched $^{28}$Si, which has almost no nuclear spins, have demonstrated extremely long $T_2^*$.[6] However, the well isolated electron spins in silicon, though offering excellent coherence times, make it difficult to use purely electrical qubit control and also limits operation speed.

Valance-band holes in silicon have even weaker hyperfine interaction, but gain a strong SOI due to their p-orbital nature. Heavy hole spins trapped in planar quantum dots for quantum information processing have been attracting significant attention since the recent theoretical predictions of long spin lifetime,[7] reduced hyperfine interaction,[8] and all electrical spin manipulation.[9] Very recently, there have been promising results obtained with single hole spins isolated in nanowires.[10–12] But the properties of holes in nanowires are very different to those of holes in devices fabricated from planar p-type MOS (pMOS) structures,[13–15] silicon-on-insulator structures,[16] or heterojunction structures.[17] The most significant difference is that in nanowires the hole ground state is the spin-1/2 light hole, whereas in two-dimensional planar structures the ground state is the spin-3/2 heavy hole. There have already been both theoretical predictions and experimental evidence that hyperfine induced spin relaxation is suppressed for heavy holes in 2D systems and quantum dots due to the suppressed hyperfine coupling to nuclei,[8,18] and that spin relaxation can be suppressed by the 2D confinement.[7,19] However, to date there has been very little work on the spin properties of holes in planar silicon devices.

A key technique for implementing spin-to-charge conversion in silicon,[20,21] and for investigating spin relaxation mechanisms,[22,23] is Pauli spin blockade.[24] For two quantum dots connected in series, if an excess spin up (down) electron has been trapped in the second dot, then ground state tunneling of electrons from the first dot onto the second dot can only occur if the additional electron has the antiparallel spin, forming a spin singlet. If the electrons in the two dots have parallel spins, forming a spin triplet, current through the double quantum dot is blocked by Pauli exclusion. This Pauli spin blockade can be lifted by spin relaxation, so measurements of Pauli spin blockade can directly probe spin relaxation mechanisms.[25,26]

Here, we present hole transport in a planar Si metal-oxide-semiconductor (MOS) based double quantum dot which offers the best of both material systems: Rapid local spin manipulation via the strong SOI of holes; and absence of hyperfine interactions. We demonstrate Pauli spin blockade in the few hole regime and identify the dominant spin relaxation mechanisms as the inter-dot tunneling coupling is varied. Upon application of a strong magnetic field, we observe an anticrossing of the singlet and triplet levels, from which we estimate the heavy hole $g$-factor and the strength of the SOI.

Figure 1(a) shows a SEM image of a multilayer gate-electrode Si MOS structure identical to the one used in this measurement. The multilayer gate structure enables great flexibility of operation from defining a high quality single dot (Supplementary Information, S1) through to defining few hole double dots, as shown schematically in Fig. 1(b), studied in this work. Lead gates (LG) are strongly negatively biased ($V_{LG} = -4$ V) to accumulate holes in source and drain reservoirs connecting to the ohmic contacts. Lower gates G1 and G2 define dot 1 and 2, and control their occupancy. Upper gate G3, in between G1 and G2, is used to control the inter-dot tunnel barrier and hence the coupling between dots. The lithographic dimensions of dot 1 and 2 are defined by the $\sim 30$ nm width of gates G1 and G2, and the $\sim 100$ width of the adjacent gate LG.[13]

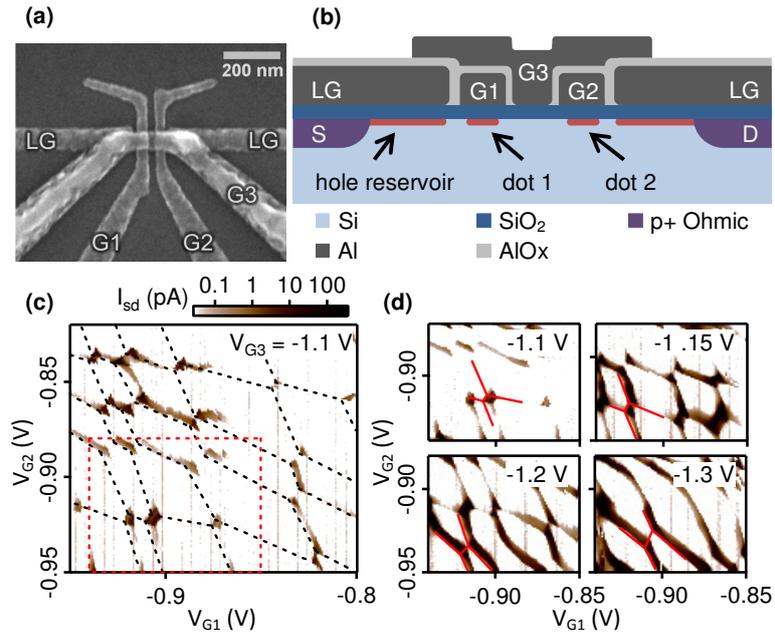

Figure 1: (a) Scanning electron microscope image of a device identical to that measured. (b) Schematic cross-section of the device when operating in double dot mode. The dot 1 (2) is defined by lower gate G1 (G2). Upper gate G3 is used to tune the inter-dot tunnel barrier. The lead gates (LG) induce source and drain reservoirs. The gate electrodes are electrically isolated from each other by thin AlO$_x$ layers. (c) Double dot charge stability diagrams with $V_{sd}$=1 mV and $V_{G3}$=-1.1 V. The source-drain current ($I_{sd}$) is measured as a function of bias on gates G1 and G2. The black dashed lines joining the triple points are superimposed on the diagram, delineating different charge configurations. (d) is a close-up of the region marked by red dotted lines in panel (c), showing how the inter-dot tunneling can be tuned with gate G3 (indicated in the upper right of each panel).

Fig. 1(c) presents charge stability diagrams of the double dot. Although we were unable to reach the last hole regime before $I_{sd}$ drops below the background noise, the large variation on the separation of the triple points suggests that we are in the few hole limit, where the orbital energy level spacing is becoming comparable to the Coulomb charging energy. Fig. 1(d) is a zoom-in of the red dashed rectangle in (c), showing how the inter-dot tunnel coupling can be tuned with gate G3. At $V_{G3} = -1.1$ V (top left of Fig. 1(d)) the tunnel coupling is weak, with isolated triple points. Making G3 more negative increases the coupling, so that the characteristic honeycomb pattern forms ($V_{G3} = -1.15$ V, top right of Fig. 1(d)). Finally at $V_{G3} = -1.3$ V (bottom right of Fig. 1(d)) the inter-dot tunnel coupling is strong, and the states in the two dots hybridize, resulting in almost diagonal lines. The increase in inter-dot coupling also causes an increase in the separation between adjacent bias triangles, as highlighted by the solid red lines.

Pauli spin blockade is shown in Figure 2. Figures 2(a) and (b) depict the (m+1,n+1) ↔ (m,n+2) inter-dot transition, while Figs. 2(c) and (d) show the (m+1,n+3) ↔ (m,n+4) inter-dot transition, here (m,n) denotes the number of holes in dot 1 and 2. Comparing data for positive and negative $V_{sd}$ we observe current rectification with suppressed transport only at positive $V_{sd}$ [Figs. 2(b,d)]. As expected, we do not observe spin blockade for the (m+1,n+2) ↔ (m,n+3) inter-dot transition (Supplementary Information S2). Spin blockade originates from spin-conserved tunneling; here we label the two relevant highest energy hole spins, or more precisely the spin-orbit doublets, by equivalent singlet or triplet states. At positive bias, the tunneling of the additional hole onto dot 2 can only happen for tunneling between S(1,1) → S(0,2) singlet states, and is blocked if the holes are in the triplet state, T(1,1). Above a certain inter-dot energy level detuning $\varepsilon$, the spin blockade can be lifted since T(0,2) can now be accessed from the T(1,1) state. Hence, as shown by the dashed trapezoids in Fig. 2(b) and (d), the spin blockade region is defined by the spacing between singlet and triplet states, which, in terms of energy spacing, is the gap between S(0,2) and T(0,2). This allows us to extract $\Delta_{ST}$ ∼800 $\mu$eV for both transitions shown in Figs. 2(b) and (d). At negative source-drain bias, the spin blockade is absent since the holes are always loaded into S(0,2) state and can then exit via S(1,1).

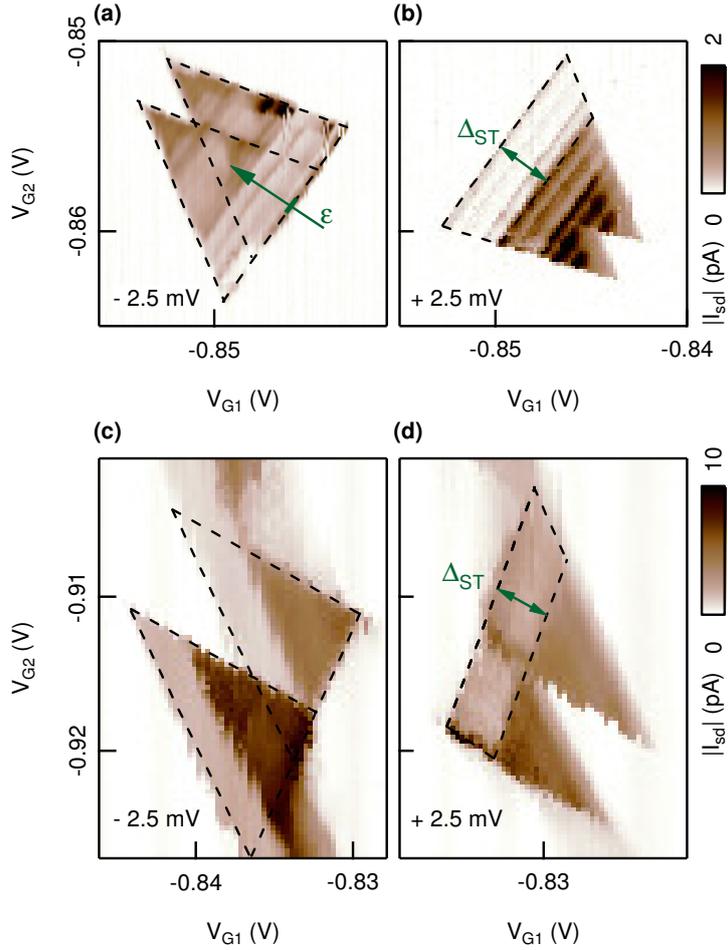

Figure 2: Bias triangles and Pauli spin blockade. $I_{sd}$ is mapped as a function of biases on gate G1 and G2 with $V_{sd}=\pm 2.5$ mV. $V_{G3}$ is set to -1.1V to give weak inter-dot coupling. (a)-(b) Hole transport through the double dot involving inter-dot transition $(m+1,n+1) \leftrightarrow (m,n+2)$. (c)-(d) Hole transport involving inter-dot transition $(m+1,n+3) \leftrightarrow (m,n+4)$. At positive bias [(b) and (d)], current is suppressed inside the singlet-triplet gap ($\Delta_{ST}$). The green arrow in (a) depicts the detuning axis direction.

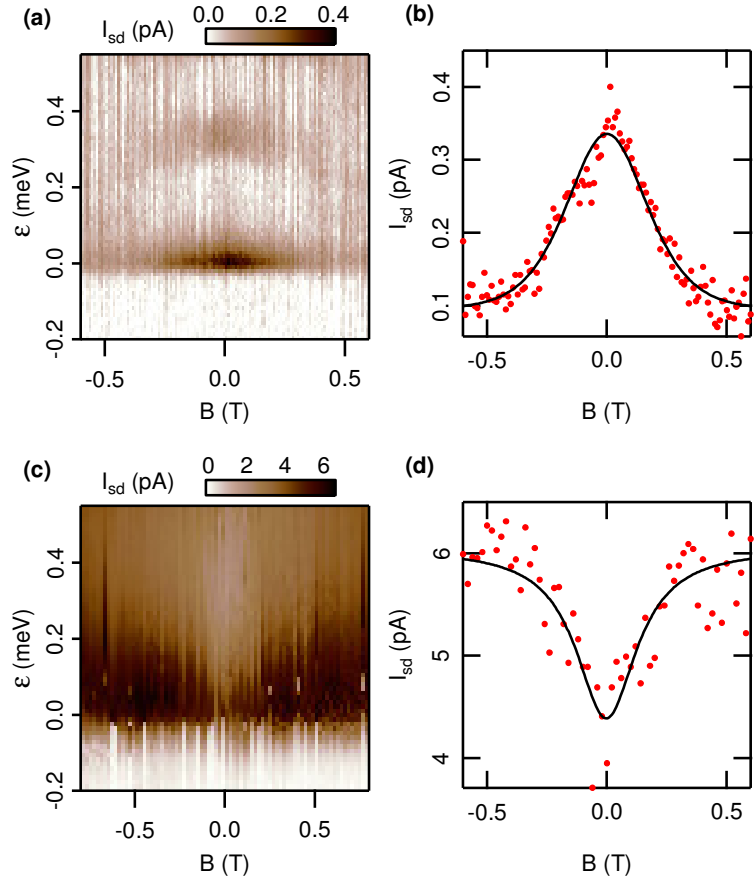

Figure 3: Double dot current in the spin blockade regime. (a) and (c) show the double dot current as a function of detuning $\varepsilon$ and out-of-plane magnetic field $B$. (a) is the same transition as in Fig. 2(b), with weak inter-dot coupling while (c) is the same transition as Fig 2(d) with two more holes in dot 2. $V_{G1}(V_{G2})$ is swept to generate the vertical traces in (a)[(c)] and then converted to $\varepsilon$. The vertical trace are shifted by aligning the $\varepsilon = 0$ transition to a straight line to eliminate occasional random charge noise offsets. (b) and (d) Line cuts at $\varepsilon = 0$ in (a) and (c). The data in (b) and (d) are fitted with models for different spin relaxation mechanisms (Supplementary Information, S4 and S5).

If the detuning is smaller than the singlet-triplet gap $\varepsilon < \Delta_{ST}$, spin blockade can also be lifted by spin relaxing mechanisms such as spin-flip co-tunneling,[23] hyperfine interactions [T(1,1) → S(1,1)],[22] or SOI [T(1,1) → S(0,2)].[22] These will allow a leakage current through the double dot when it is in the spin blockade regime. To elucidate the leading spin relaxation mechanism, the leakage current is studied as a function of detuning $\varepsilon$ and out-of-plane magnetic field $B$. Figure 3(a) maps the leakage current in the spin blockade region in Fig. 2(b). Fig. 3(b) shows a line cut around $\varepsilon = 0$ in Fig. 3(a), revealing a peak in the source-drain current around B=0. The width of the peak, $\sim 400$ mT, excludes the hyperfine interaction as the dominant spin relaxation mechanism.[27] Instead the peak is well described by spin-flip co-tunneling in the regime of weak inter-dot coupling and weak local dephasing[26](Supplementary Information S3). From fitting our data to the theory of ref. 26, we can extract the hole temperature as $\sim$170 mK and the tunneling rate to the source and drain as $\sim$39 $\mu$eV. These values are consistent with similar data from almost identical electron double dot devices,[23] and support the interpretation that the double dot is in the weak inter-dot coupling regime. This data also allows us to extract the spin relaxation rate due to SOI at higher magnetic field, where the spin-flip co-tunneling is suppressed. The finite spin blockade leakage current at higher $B$ ($B > 0.5$T in Fig. 3(a)) is due to the SOI induced relaxation rate $\Gamma^{SO}$, so that we can extract $\Gamma^{SO} \sim 2.5$ neV, which is consistent with the value of $\sim 4.6$ neV extracted from inter-dot tunneling coupling[25] (Supplementary Information S3).

To further resolve the influence of SOI on inter-dot spin relaxation, we need stronger inter-dot coupling.[25] Increased inter-dot tunneling coupling is achieved by introducing two more holes in dot 2, which, by comparing Fig. 2(d) to (b), shows up as further separated triple points, higher off-resonance current, and a greater extent of the co-tunneling current beyond the bias triangles. The spin blockade leakage current is mapped for the (m+1,n+3) $\rightarrow$ (m,n+4) transition in Fig. 3(c). Fig. 3(d) shows a line cut around $\varepsilon = 0$ in Fig. 3(c). By increasing the inter-dot tunnel coupling, the peak in $I_{sd}$ around $B = 0$ T in Fig. 3(b) becomes a dip in Fig. 3(d). The suppression of leakage current at $B = 0$ and the rapid lifting of spin blockade with finite field signifies strong SOI induced hybridization of the T(1,1) and S(0,2) states. The leakage current dip can be well fitted by a Lorentzian, in good agreement with the transport model assuming both strong SOI and strong inter-dot coupling[25](Supplementary Information S4). In this regime, the leakage current is limited by the spin relaxation rate between (1,1) states, and our fitting yields a relaxation rate of the order 3 MHz, comparable to the values reported for electrons in InAs nanowire double dot.[22] Since the relaxation rate between (1,1) states is not directly related to the inter-dot tunneling, the reduced lifetime caused by adding two more holes may be due to an enhancement of the SOI, which is highly sensitive to the size of the dot.

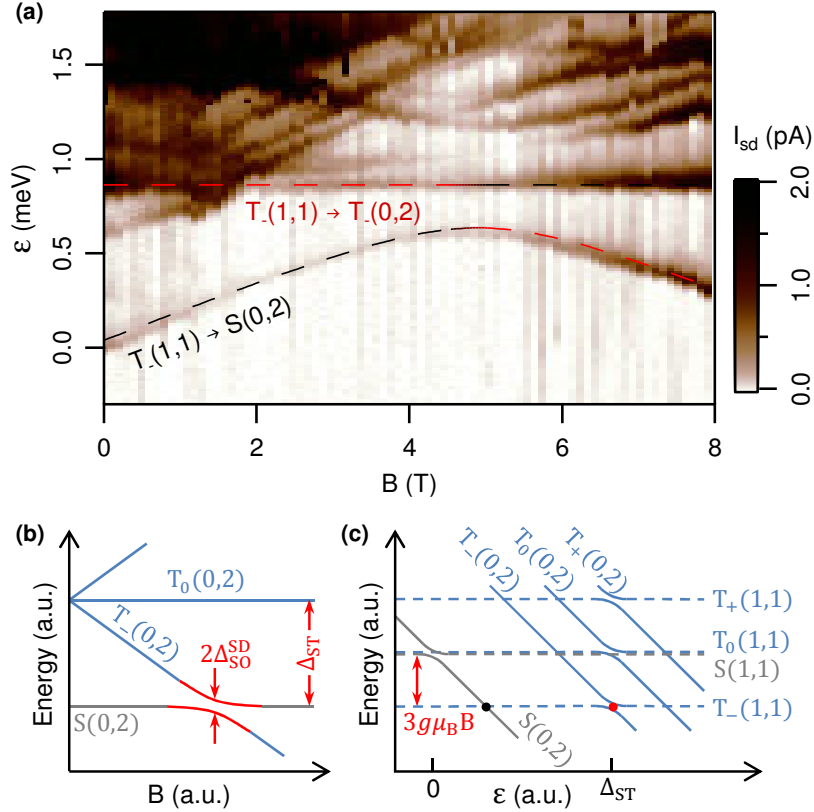

Figure 4: (a) Double dot current as a function of out-of-plane magnetic field and inter-dot energy detuning. The map is generated in the same way as Fig. 3(a), but over a larger $\varepsilon$ and $B$ range. Below $B = 4$ T, the bottom two lines are attributed to $T_-(1,1) \to S(0,2)$ and $T_-(1,1) \to T_-(0,2)$ transitions. To eliminate occasional offsets due to charge noise, traces have been shifted vertically by aligning the second resonance transition to a straight line (Supplementary Information S6). (b) Energy levels of the double dot in the (0,2) configuration. When the Zeeman energy approaches $\Delta_{ST}$ ($3g\mu_B B \sim \Delta_{ST}$), the SOI mixes $S(0,2)$ with $T_-(0,2)$ so that they anticross. (c) Double dot energy diagram as a function of inter-dot energy detuning $\varepsilon$ at finite magentic field. The magnetic field lifts the degeneracy of the triplets $T_+$, $T_0$, and $T_-$. The black dot marks the spin non conserved tunneling with the mixing of $T_-(1,1)$ and $S(0,2)$ by SOI. The red dot marks the conserved tunneling from $T_-(1,1) \to T_-(0,2)$.

The mixing of intra-dot spin states by SOI can be revealed at higher field.[28] Figure 4(a) shows a map of the double dot current in the same manner as Fig. 3(a) over a larger range of detuning and magnetic field. The lines in the map are resonance current peaks corresponding to alignment of energy states between dot 1 and dot 2. Since the inter-dot coupling is weak, and the $g$-factor is negative,[29] the bottom two lines are ascribed to the $T_-(1,1) \to S(0,2)$ and $T_-(1,1) \to T_-(0,2)$ transitions as sketched in Figs. 4(b) and (c). The $T_-(1,1) \to S(0,2)$ transition, as marked by black dot in Figs. 4(c), moves to higher detuning with increased $B$ since $T_-(1,1)$ moves to lower energy. The $T_-(1,1) \to T_-(0,2)$ transition, as marked by red dot in Figs. 4(c), does not move in detuning when $B$ is increased since there is no change in spin number.

At $B \sim 5$ T, we observe an avoided crossing between the bottom two lines in Fig. 4(a), and an increase in current amplitude for the lower transition for B> 5 T. As shown in Fig. 4(b), $T_-(0,2)$ moves closer to $S(0,2)$ with increased $B$. When the Zeeman energy approaches $\Delta_{ST}$, $T_-(0,2)$ anticrosses with $S(0,2)$ due to SOI. If we ascribe the gap in Fig. 4(a) to the SOI induced repulsion between $S(0,2)$ and $T_-(0,2)$, we can fit the the bottom line to the standard anticrossing gap expression.[30] From this fitting, we extract the hole Landé $g$-factor as 0.93 and the single dot spin-orbit gap $\Delta_{SO}^{SD}$ of 110 $\mu$eV. The extracted heavy hole $g$-factor is similar to the theoretical prediction of 0.84 for 2D heavy holes,[29] and the values measured in boron acceptors in Si.[31] Similar values to the extracted $\Delta_{SO}^{SD}$ have been reported in InAs[30] and InSb nanowires.[28] The spin orbit length can be estimated to be $\sim 110$ nm,[30] which is considerably larger than the size of dot 2. However, we note that some caution should be exercised since the two lines identified in Fig. 4(a) do not follow the conventional evolution of a standard anticrossing,[28,30] and there are several resonances at higher $\varepsilon$ which cannot be assigned. In addition the dependence of these two transitions on in-plane magnetic field (Supplementary Information S6) is not well understood, considering the complex spin properties of spin-3/2 holes.

To summarize, we have studied spin dependent transport in a hole double dot system based on a highly flexible multi-gate silicon pMOS structure. We observe Pauli spin blockade at different charge configurations, and observe a leakage current due to spin relaxation. By studying the magnetic field dependence of the leakage current, we identify spin-flip cotunneling as the dominant spin relaxation mechanism at weak inter-dot coupling, while spin-orbit driven spin relaxation dominates at stronger inter-dot coupling. Application of a strong magnetic field reduces the singlet-triplet splitting, from which we obtain a hole $g$-factor consistent with spin-3/2 heavy holes. An avoided crossing is observed, from which we estimate the spin-orbit strength. These results give an insight into the properties of heavy hole spins in silicon quantum dots and provide a pathway towards spin-based qubits and quantum information processing using heavy holes.

## Methods

**Device fabrication.** The device studied in this work is fabricated from a high-resistivity ($\rho > 10$ k$\Omega$-cm) natural (001) silicon substrate. The p+ ohmic regions in Fig. 1(b) are prepared by solid source boron diffusion at $\sim 975$ °C with a peak doping density of $\sim 2 \times 10^{19}$/cm$^3$. Subsequently, a 5.9 nm gate dielectric (SiO$_2$) is grown by dry oxidation in the active region of the device. Lower gates G1 and G2, lead gate LG, and upper gate G3 are sequentially patterned on the gate dielectric by electron beam lithography, thermal evaporation of aluminum, and metal lift-off. Between each patterning, the lower aluminum electrodes are oxidized for 10 min at $\sim 150$ °C in air to form $\sim 4$ nm AlO$_x$ for electrical isolation. The final stage is forming gas (95% N$_2$/5% H$_2$) annealing to reduce the Si/SiO$_2$ interface disorder and enhance low-temperature performance.

**Experimental setup.** Electrical transport measurements are carried out in a dilution fridge with a base temperature lower than 30 mK. All signal lines are cold filtered.

# Acknowledgement

The authors thank W. A. Coish and D. Culcer for many helpful discussions, A. M. See, O. Klochan, L. A. Yeoh and A. Srinivasan for help with the dilution refrigerator, and J. Cochrane for technical support. A.R.H. acknowledges support from the Australian Research Council (DP120102888, DP120101859 and DP150100237). F.E.H. and A.S.D. acknowledge support from the Australian Research Council (CE110001027) and the U.S. Army Research Office (W911NF-13-1-0024). Devices for this study were fabricated with support from the Australian National Fabrication Facility node at UNSW.

# Supporting Information Available

Further information on single quantum dot operation, charge stability diagram of double quantum dot, lever-arm of lower gate G1 and G2, fitting equations to the leakage current in different inter-dot tunnel coupling, and double dot current as a function of in-plane magnetic field is presented.

This material is available free of charge via the Internet at `http://pubs.acs.org/`.

# Graphical TOC Entry

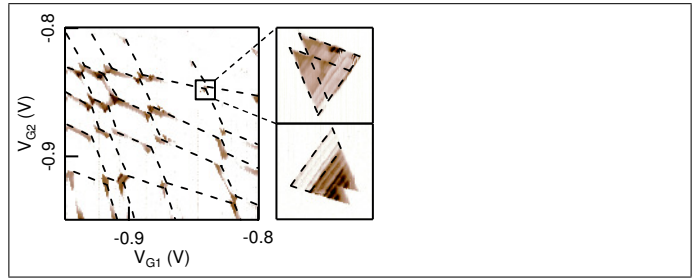

# Supplementary Information:
# Pauli Spin Blockade of Heavy Holes in a Silicon Double Quantum Dot

## S1. Single hole transistors with quantum dots defined under different gate electrodes

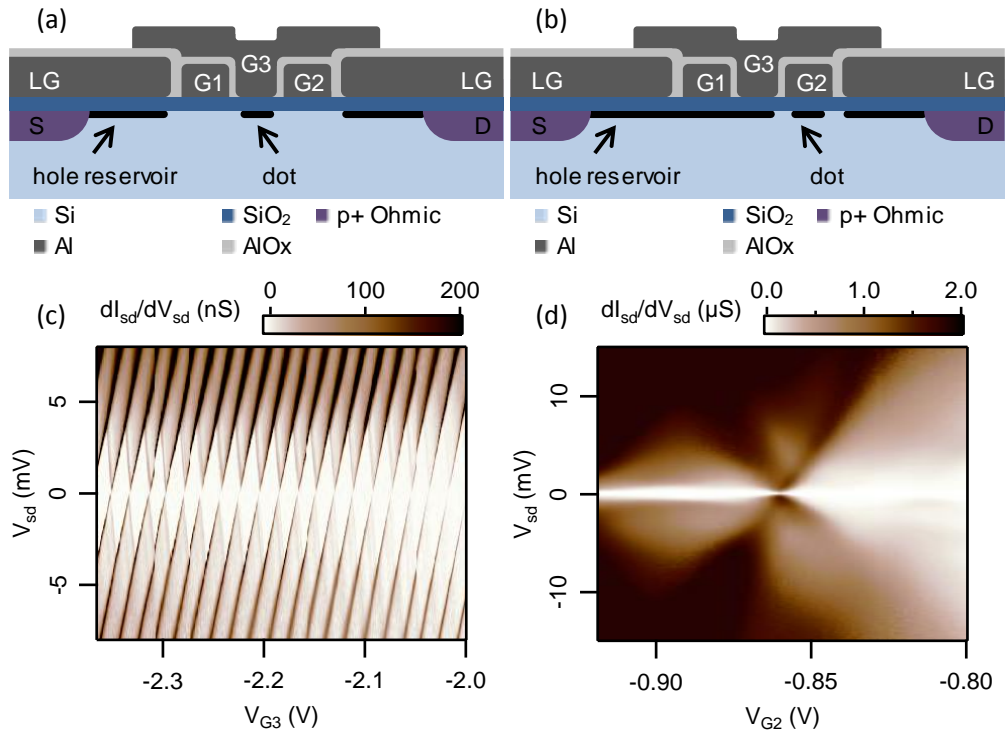

Figure S1: (a) and (b) Schematic cross section of the Si MOS multi-gate structure operating as a single hole transistor. For (a), a single quantum dot is induced by upper gate G3. Biases on finger gates G1 and G2 are set to $V_{G1}$=-0.8V and $V_{G2}$=-0.83V. For (b), a single quantum dot is defined by G2 with $V_{G1}$=-1.4V and $V_{G3}$=-2V. Lead gates are set to -4V. (c) and (d) Source-drain bias spectroscopy of the quantum dots depicted in panels (a) and (b). For (c), the dot is in many hole regime with weak coupling to source and drain reservoirs. For (d), the dot is strongly coupled to the reservoirs with pronounced co-tunneling current in Coulomb blockade regions.

Figure S1 shows that with different gate bias configuration, the multi-gate structure can operate as a single hole transistor (SHT) with isolated hole quantum dots defined under different gates. As shown in Fig. S1(a), by setting G1 and G2 to depleting mode and G3 to accumulation mode, a single quantum dot can be defined under G3 and operate in a similar way as in Ref. S1. Bias spectroscopy for this configuration is shown in Fig. S1(c). The large number of highly periodic Coulomb diamonds demonstrates that the dot is in many hole regime and that the Si MOS multi-gate structure is highly stable.

Fig. S1(b) shows that a single quantum dot can be defined under G2 by setting G1 and G3 to accumulation mode. As shown in Fig. S1(d), there is a finite current inside Coulomb diamonds, suggesting a high transparency of the source and drain tunneling barriers. This is consistent with the geometry of the structure, since the width of the tunneling barrier is defined by the ~4nm AlO$_x$ gap between Al gate electrodes. Inside the Coulomb diamond in Fig. S1(d), with $V_{G3}$ around -0.9V, an enhanced conductance can be found when the source-drain bias ($V_{sd}$) is higher than ~0.6mV. This is corresponding to inelastic co-tunneling via an excited state [S2] and an orbital level spacing of ~0.6meV can be extracted. This value is consistent with the singlet-triplet spacing extracted from double dot measurements with a similar plunger gate bias.

## S2. Proof of Pauli spin blockade: double dot charge stability diagram and bias triangles

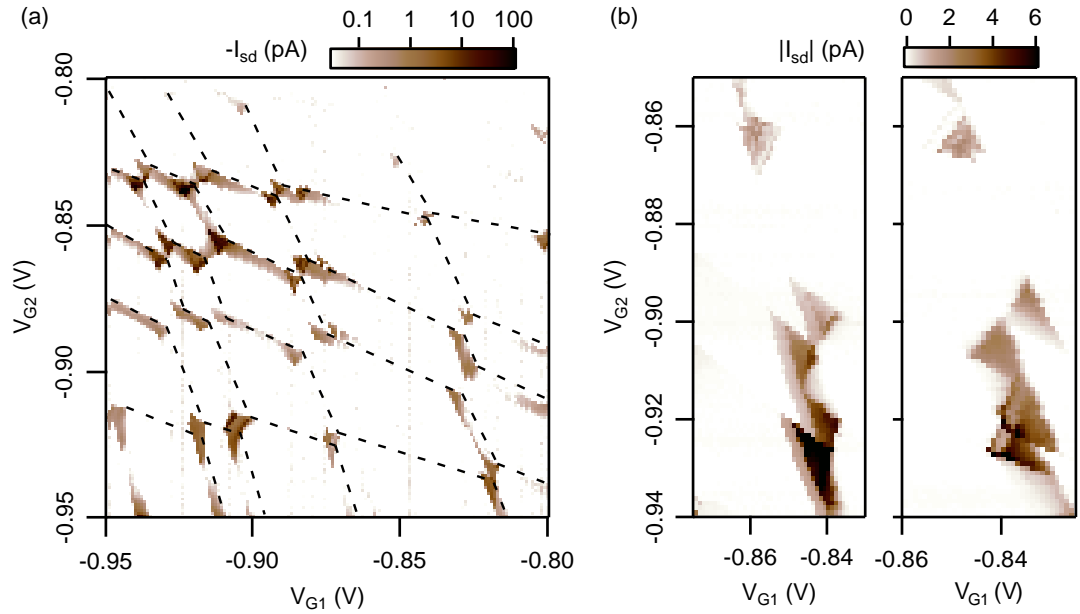

Figure S2: (a) Double dot charge stability diagram with upper gate G3 set to -1.1V and source-drain bias set to -1V. (b) Detailed gate G1 vs. G2 map showing inter-dot transitions of (m+1,n+1)↔(m,n+2), (m+1,n+2)↔(m,n+3), and (m+1,n+3)↔(m,n+4). The source-drain bias is set to -3mV (left) and 3mV (right).

Figure S2(a) shows the double dot charge stability diagram with weak inter-dot coupling. The dotted lines joint the triple points. At the top right corner of Fig S2(a) the number of holes is small and the source-drain current drops below the background noise so we cannot determine the exact hole occupancy. However, even for adjacent charge configurations, there are large variations of the size of the parallelograms in this regime, suggesting that we are approaching the last few holes in the double dot system.

To show that the double dot system exhibits Pauli spin blockade we compare the pairs of bias triangles with different hole occupation in dot 2. From top to bottom, both maps in Fig. S2(b) shows three pairs of bias triangles with consecutive inter-dot transitions of (m+1,n+1)↔(m,n+2),

(m+1,n+2)↔(m,n+3), and (m+1,n+3)↔(m,n+4). Comparing with the negative source-drain bias triangles, we see that at positive bias, there is a suppression of the current at the base in the bias triangles for the top and bottom pairs in Fig. S2(b). In contrast, by changing the hole occupancy on dot 2 by one, there is no suppression of the current for the middle triangle pair. Though the spin filling of hole quantum dot can be highly polarized [S3,S4], the alternating appearance of suppression current is a characteristic signature of ever-odd spin filling and Pauli spin blockade [S5].

## S3. Lever-arm of the double dot

Information about the dot size and few hole limit can be obtained from the geometry of the bias triangles in Figure 2 in the main text. By the ratio of the plunger gate voltage span to $V_{sd}$, the gate voltage to dot energy conversion ratio, or lever-arm $\alpha$, of G1 and G2, can be calculated as shown in Table S1. If we follow the semi-classical interpretation [S6], $\alpha_{1(2)} = C_{G1(G2)}/C_{total}$, which is the ratio of capacitive coupling of G1(G2) to the total environment of the dot 1 (dot 2). Since dot 1 and 2 have identical lithographic dimensions, the fact that $\alpha_1 > \alpha_2$ suggests dot 1 has smaller coupling to the other gates and reservoirs, and hence the dot 1 size is smaller. Meanwhile, when two move holes are introduced into dot 2, $\alpha_2$ exhibits a ~25% decrease, implying a either considerable expansion in the size of dot 2, a drastically altered orbital wavefunction [S7], or the electrostatic shift of orbital level [S8], all of which further confirms the double dot is approaching the last few holes.

Table S1: Lever-arm $\alpha$ of G1 and G2 for different numbers of holes on the double dot

|  | (m+1,n+1)→(m,n+2) | (m+1,n+3)→(m,n+4) |
|---|---|---|
| $\alpha_1$ | 0.32 | 0.29 |
| $\alpha_2$ | 0.29 | 0.21 |

## S4. Spin flip co-tunneling in spin blockade regime

The rate equation describing low spin-flip co-tunneling in weak inter-dot coupling regime is [S9]:

$$I(B, \varepsilon = 0) = \frac{4}{3} e c k_B T \frac{3g\mu_B B/k_B T}{\sinh(3g\mu_B B/k_B T)} + I_B \quad (S1)$$

with the co-tunneling coefficient

$$c = \frac{h}{\pi}\left[\left(\frac{\Gamma_L}{\Delta}\right)^2 + \left(\frac{\Gamma_R}{\Delta - 2U' - 2e|V_{sd}|}\right)^2\right] \quad (S2)$$

where $e$ is the electron charge, $k_B$ is the Boltzmann constant, $T$ is the hole temperature, $g$ is the Landé g-factor of spin-3/2 heavy holes (0.93 is used, Zeeman splitting $\Delta E_Z = 3g\mu_B B$), $\mu_B$ is the Bohr magneton, $B$ is the magnetic field, $I_B$ is the background current offset, $h$ is the Planck constant, $\Gamma_L$ ($\Gamma_R$) is the tunneling rate of the source (drain) to dot 1 (2), $\Delta$ is the depth of the two-hole level [S9], and $U'$ is the inter-dot charging energy.

As shown in Figure 2(b), there is no observable change of the current amplitude along the zero detuning line, so we assume $\Gamma_L \sim \Gamma_R = \Gamma$ and $\Delta \sim U' \sim 1.25$meV [S11]. Then the fitting shown in Figure 3(b) yields a hole temperature $T \sim 170$ mK and a tunneling rate to source/drain $\Gamma \sim 39$ μeV.

The tunneling through the double dot in the absence of spin blockade can be described by the equation [S6]:

$$I(\varepsilon) = e \frac{\Gamma_3 t^2}{\left(\frac{\varepsilon}{\hbar}\right)^2 + \frac{\Gamma_3^2}{4} + t^2\left(2 + \frac{\Gamma_3}{\Gamma_1}\right)} \tag{S3}$$

where $\varepsilon$ is the double dot energy detuning, $\Gamma_1$ ($\Gamma_3$) is the rate of a hole entering (exiting) the double dot, and $t$ is the inter-dot tunneling rate. Equation S3 can be used to describe the same inter-dot transition with negative source-drain bias, i.e. no spin blockade, which gives $\Gamma_1 = \Gamma_R$ and $\Gamma_3 = \Gamma_L$. By assuming $\Gamma_L \sim \Gamma_R = \Gamma$, the fitting result of Eq. (S3) to $I_{sd}$ is plotted in Figure S3. The asymmetry of the current peak around ε=0 is due to inelastic tunneling processes at positive detuning so only the data points at the rising edge of the peak are used. $\Gamma \sim 68$ μeV and $t \sim 0.56$ μeV can be extracted from the fitting.

The tunneling rates to source/drain reservoir $\Gamma$ from two different fitting methods agree well with each other (fitting to spin flip co-tunneling at positive V$_{sd}$ gives $\Gamma \sim 39$ μeV and fitting to elastic inter-dot tunneling at negative V$_{sd}$ gives $\Gamma \sim 68$ μeV). Furthermore, the extracted values are consistent with the limits of weak inter-dot tunneling and weak co-tunneling ($\sqrt{2}t < k_BT, ck_BT \ll \Gamma$), justifying the use of Eq. (S1) in analyzing the leakage current.

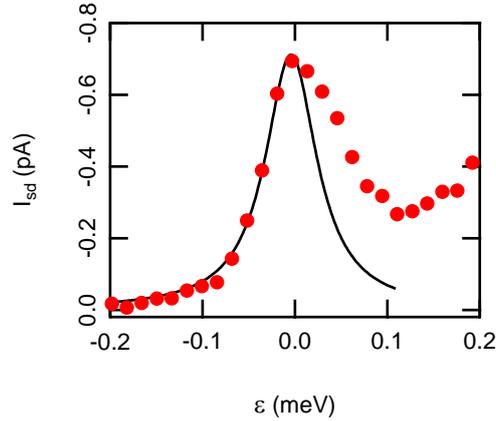

Figure S3: (m,n+2)→(m+1,n+1) inter-dot transition (non-blockaded, V$_{sd}$=-2.5mV) around zero detuning. The black solid line shows fitting result of Eq. (S3) to the ε<0 data points.

While fitting the spin-flip co-tunneling leakage current, a background current $I_B$ = 95 fA can be extracted from the fitting. This current continues out to higher magnetic field (with reduced amplitude) as shown in Figure 4, and is attributed to spin orbit interaction induce leakage current. By a rough estimation assuming that the spin orbit induced singlet-triplet coupling $\Gamma^{SO}$ is the only rate limiting factor in $I_B$, we can use the relation of $e\Gamma^{SO} \sim I_B$ to estimate $\Gamma^{SO} \sim 2.5$ neV.

According to Ref. [S13], the spin orbit induced relaxation rate can also be estimated based on inter-dot coupling as $\Gamma^{SO} \sim t^2/\Gamma$ where $\Gamma$ and $t$ are extracted from the fitting of Eq. (S3). This alternative analysis yields $\Gamma^{SO} \sim 4.6$ neV, confirming the ascription of $I_B$ to spin orbit interaction.

## S5. SOI induced leakage current in the spin blockade regime

The leakage current at zero detuning due to spin orbit interaction can be described by a Lorentzian line shaped dip around zero magnetic field [S13]:

$$I = I_{max}\left(1 - \frac{8}{9}\frac{B_c^2}{B^2 + B_c^2}\right) + I_B \tag{S4}$$

with dip width parameter $B_c$ and dip height $I_{max} = 4e\Gamma_{rel}$, where $\Gamma_{rel}$ is the spin relaxation rate. As show in Fig. 3(d), by fitting the source-drain current in the regime of spin blockade and increased inter-dot tunneling coupling to Eq. (S4), we can extract $\Gamma_{rel} \sim 3$ MHz.

The background current $I_B$ could be attributed to higher order co-tunneling processes [S6].

## S6. Double dot current as a function of in-plane magnetic field and inter-dot energy detuning

The (0,2) spectrum with in-plane magnetic field is plotted in Figure S4(b). This map differs greatly from Figure 4(a) in main text and is not yet fully understood. For in-plane field, the resonance lines are considerably wider, which blurs the T-(1,1)→T-(0,2) transition. Extra lines within the singlet-triplet gap can be found, which has also been reported in InAs self-assembled quantum dot [S14]. However, unlike Ref. S14 where the extra state does not interact with the states of interest, a weak anti-crossing can be traced around 2T between the T-(1,1)→S(0,2) and the unidentified state. A pronounced crossing around 4T can also be found. However, the identification of its origination is challenging without the knowledge of the exact number of holes in each dot, and the energy spectrum of holes in 2D systems is not fully understood.

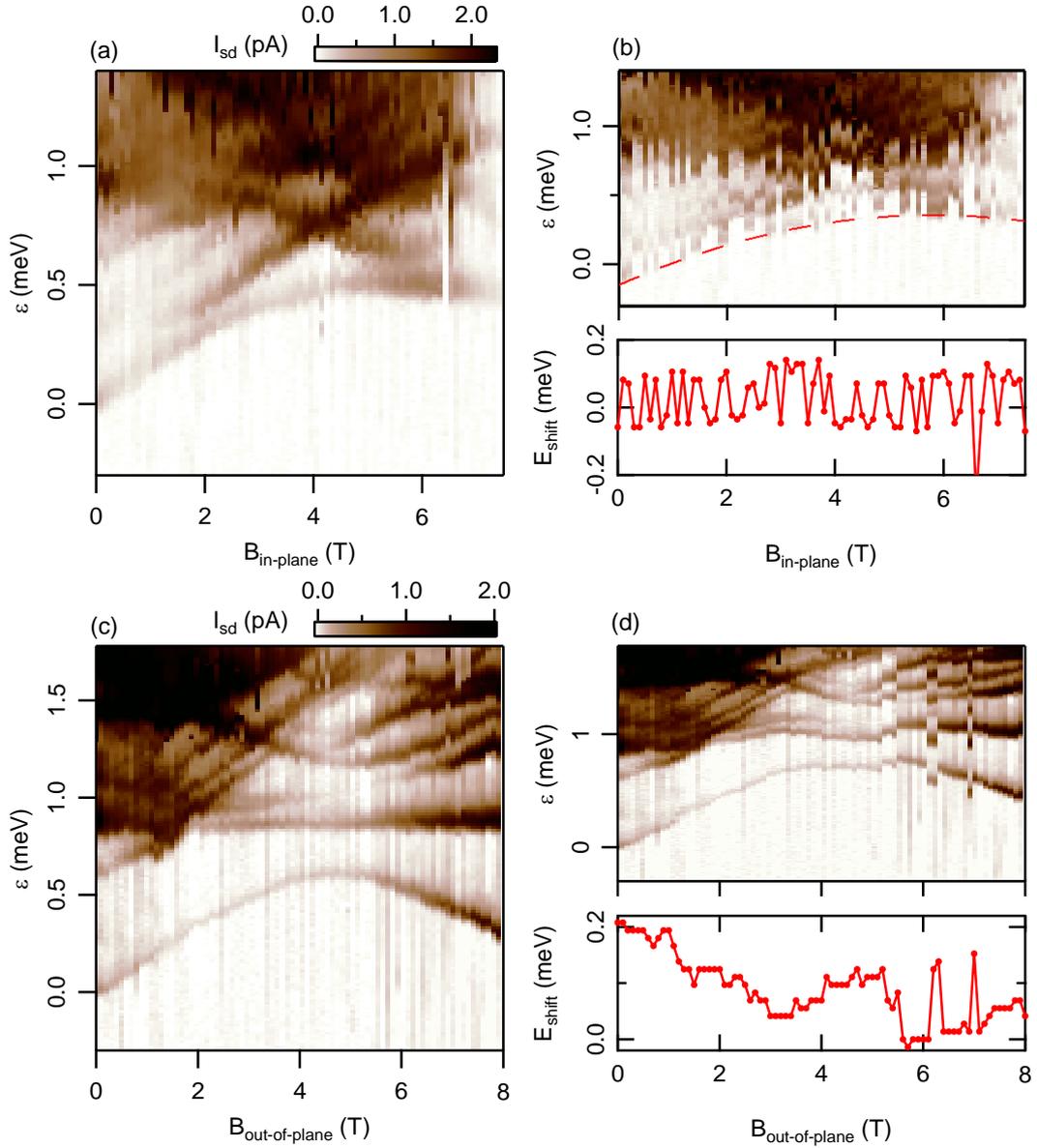

Figure S4: (a) Double dot current as a function of detuning and in-plane magnetic field. The map was taken over a 15-hour period, as gate bias was swept for many different magnetic fields. To eliminate the effect of random telegraph noise during the measurement, we fit a parabola to the bottom transition in the raw data as marked by the red dashed line in the top of panel (b). The offset applied to each sweep relative to the parabola is plotted in the bottom of panel (b), where the two-level fluctuator can clearly be seen. (c) Double dot current as a function of detuning and out-of-plane magnetic field. The map was the same as Figure 4(a) in the main text, which was taken over a 16-hour period. The device stability was much better during this measurement, but charge noise was still present as shown in the raw data in the top of panel (d). Since there are several holes in the double dot, and the g-factor may differ in the two dots, for simplicity we align the second transition line to a horizontal line. The offset applied to each sweep is plotted in the bottom of panel (d). We emphasize that this adjustment does not change the energy gap extracted from the data, since the spacing between transitions remains unchanged.